%Paper: hep-ph/9206222
%From: golden@weyl.bu.edu
%Date: Mon, 15 Jun 92 15:18:48 -0400
%Date (revised): Tue, 16 Jun 92 11:32:46 -0400

%This document uses the Harvard macro package harvmac.tex
%There are three postscript files for the three figures appended to this file
%Look for the words "CUT HERE"

\input harvmac
\noblackbox
\def\lae{\raise-.5ex\vbox{\hbox{$\; <\;$}\vskip-2.9ex\hbox{$\; \sim\;$}}}
\def\gae{\raise-.5ex\vbox{\hbox{$\; >\;$}\vskip-2.9ex\hbox{$\; \sim\;$}}}
\def\slash#1{\raise.15ex\hbox{/}\kern-.57em #1}

\def\ie{{\it i.e.}}

\def\two#1{\raise1.35ex\hbox{$\leftrightarrow$}\kern-.88em#1}
\def\lefta#1{\raise1.35ex\hbox{$\leftarrow$}\kern-.61em#1}
\def\righta#1{\raise1.35ex\hbox{$\rightarrow$}\kern-.61em#1}
\def\Dslash{\raise.15ex\hbox{/}\kern-.77em D}
\def\np#1#2#3{Nucl. Phys. {\bf #1} (#2) #3}
\def\pl#1#2#3{Phys. Lett. {\bf #1} (#2) #3}
\def\prl#1#2#3{Phys. Rev. Lett. {\bf #1} (#2) #3}
\def\pr#1#2#3{Phys. Rev. {\bf #1} (#2) #3}
\def\prd#1#2#3{Phys. Rev. D {\bf #1} (#2) #3}
\def\vsl{\raise.15ex\hbox{/}\kern-.57em v}
\def\ie{{\it i.e.}}

\def\etal{{\it et al}}
\def\tr{{\rm tr}}
\def\Lt{{{\cal L}_2}}
\def\Lf{{{\cal L}_4}}
\def\spi{{1 \over 16\pi^2}}
\def\lgmm{\spi \log {m^2\over\mu^2}}
\def\DD{\lgmm}

\Title{\vbox{\hbox{HUTP--92/A025}\hbox{BUHEP--92--18}\hbox{hep-ph/9206222}%
}}{ \vbox{\centerline{Analyticity, Crossing Symmetry and the}
\vskip2pt\centerline{Limits of Chiral Perturbation Theory}}}

%For more complicated situations, substitute for {\it either\/} argument:
%\Title{\vbox{\baselineskip12pt\hbox{HUTP-88/A000}\hbox{SLAC-PUB 88-001}
%               \hbox{photocopy at own risk}}}
%{\vbox{\centerline{This title is too long to fit}
%       \vskip2pt\centerline{comfortably on one line*}}}
%   \footnote{}{*optional footnote on title}

\centerline{R. Sekhar Chivukula$^{a,1}$}
\bigskip\centerline{Michael J. Dugan$^{a,b,2}$}
\bigskip\centerline{\it and}
\bigskip\centerline{Mitchell Golden$^{a,3}$}
\footnote{}{$^a$Boston University, Department of Physics, 590 Commonwealth
Avenue, Boston, MA 02215}
\footnote{}{$^b$Lyman Laboratory of Physics,
Harvard University, Cambridge, MA 02138}
\footnote{}{$^1$SEKHAR@WEYL.BU.EDU}
\footnote{}{$^2$DUGAN@HUHEPL.BITNET}
\footnote{}{$^3$GOLDEN@WEYL.BU.EDU}

%if too many authors for abstract on same page, say   \vfill\eject\pageno0

\vskip .2in

The chiral Lagrangian for Goldstone boson scattering is a power series
expansion in numbers of derivatives.  Each successive term is suppressed by
powers of a scale, $\Lambda_\chi$, which must be less than of order $4\pi
f/\sqrt{N}$ where $f$ is the Goldstone boson decay constant and $N$ is the
number of flavors. The chiral expansion therefore breaks down at or below $4
\pi f/\sqrt{N}$.  We argue that the breakdown of the chiral expansion is
associated with the appearance of physical states other than Goldstone bosons.
Because of crossing symmetry, some ``isospin'' channels will deviate from
their low energy behavior well before they approach the scale at which their
low energy amplitudes would violate unitarity.  We argue that the estimates of
``oblique'' corrections from technicolor obtained by scaling from QCD are
untrustworthy.

\Date{6/92} %replace this line by \draft  for preliminary versions
             %or specify \draftmode at some point
%\draft

\newsec{Introduction}

The chiral Lagrangian \ref\CCWZ{S. Weinberg, \pr {166} {1968} {1568}\semi S.
Coleman, J. Wess, and B. Zumino, \pr {177} {1969} {2239} \semi C. Callan, S.
Coleman, J. Wess, and B. Zumino \pr {177} {1969} {2246}.} is a compact way of
calculating the amplitudes for low-energy processes involving Goldstone
bosons.  Though it was originally developed to describe pions, it has more
recently been applied to calculate amplitudes for processes involving the
longitudinal components of the $W$ and $Z$ gauge bosons \ref\ewchiral{T.
Appelquist and C. Bernard, \prd {22} {1980} {200} and \prd {23} {1981}
{425}\semi A.~Longhitano, \prd {22} {1980} {1166} and \np {B231} {1984}
{205}.} \ref\chanowitz{ M.~S.~Chanowitz and M.~K.~Gaillard, \np {B261} {1985}
{379}.} \ref\golden{M. S.  Chanowitz, M. Golden, and H.  Georgi, \prl {57}
{1986} {2344}, and \prd {35} {1987} {1490}.}.  The equivalence theorem
\chanowitz\ insures that, at energies large compared to $M_W$, the scattering
amplitudes of longitudinally polarized gauge bosons are approximately the same
as those of the Goldstone bosons which would be present in the ungauged
theory. In technicolor models \ref\techni{S.  Weinberg, \prd {19} {1979}
{1277}\semi L. Susskind, \prd {20} {1979} {2619}.}, frequently there are also
additional, approximate Goldstone bosons \ref\el{E. Eichten and K. Lane, \pl
{90B} {1980} {125}\semi S. Dimopoulos and L. Susskind, \np {B155} {1979}
{237}.}.  For example, the one-family model \ref\onefamily{E. Farhi and L.
Susskind, \prd {20} {1979} {3404}.} has 60 pseudo-Goldstone bosons, in
addition to the 3 ``swallowed'' degrees of freedom.  In this case, the chiral
Lagrangian may also be used to describe low-energy processes involving
pseudo-Goldstone bosons as well as longitudinal $W$ or $Z$ bosons
\ref\baggerherr{J. Bagger, S. Dawson, and G.  Valencia, \prl {67} {1991}
{2256}.}.

Consider a model in which the symmetry breaking pattern is $SU(N)_L \times
SU(N)_R \to SU(N)_V$ ($N=2$ in the simplest technicolor model and $N=8$ in
the one-family model). The most general chirally invariant Lagrangian \CCWZ\
may be written in terms of the field
\eqn\exppi{
\Sigma = \exp(2 i \pi^a T^a / f)~,
}
where the $\pi^a$ (which we refer to as ``pions'') are the Goldstone boson
fields, the $T^a$ are the generators of $SU(N)$, normalized to $\tr T^aT^b =
\delta^{ab}/2$, and $f$ is the analog of the pion decay constant. Under
a chiral transformation, the field $\Sigma$ transforms as $\Sigma \to L \Sigma
R^\dagger$, with $L \in SU(N)_L$ and $R \in SU(N)_R$.  The most general
chirally invariant Lagrangian can be written as an expansion in powers of
derivatives.  There are no nontrivial chirally invariant terms involving no
derivatives and only
\eqn\lag{
\Lt^{(0)} = {f^2 \over 4}\tr \partial^\mu \Sigma^\dagger
\partial_\mu \Sigma~
}
with two derivatives. Additional terms with more derivatives are suppressed
by powers of some momentum scale, denoted $\Lambda_\chi$.

Chiral perturbation theory is an expansion in $p^2/\Lambda^2_\chi$.
The utility of the chiral Lagrangian arises from the fact that at energies
less than $\Lambda_\chi$, the interactions of exact Goldstone bosons are
determined by $\Lt^{(0)}$, {\it i.e.} they are entirely determined by the
symmetry structure of the theory. For this reason, these lowest order
predictions are universal \golden.

At energies near or above $\Lambda_\chi$, however, all terms in the expansion
contribute and the chiral Lagrangian becomes effectively useless.  Here the
amplitudes become model dependent. The expansion must fail at or before an
energy scale associated with the appearance of additional particles or
resonances: for example, in the case of the one Higgs-doublet standard model
the chiral expansion fails at an energy scale of order the Higgs mass. In a
weakly coupled theory, these particles will generally be light and the chiral
expansion will fail at low energies.

In a strongly interacting theory, such as QCD or technicolor, the resonances
will generally be heavy, and the predictions of chiral perturbation theory may
be used to fairly high energies. It is important, therefore, to understand how
large $\Lambda_\chi$ may be. This question was first addressed by Weinberg
\ref\wein{S. Weinberg, Physica {\bf 96A} (1979) 327\semi see also H. Georgi and
A. Manohar, \np {B234} {1984} {189}.} who argued that, since the higher order
terms are required as counterterms to loops involving the lowest order
interactions, it was inconsistent to assume that the size of these terms was
smaller than that typical of the corresponding loop correction. This leads to
the estimate of naive dimensional analysis that $\Lambda_\chi$ must be less
than or about $4\pi f$.

In this paper, we discuss how the bound on $\Lambda_\chi$ varies with $N$,
\ie\ how it depends on the number of Goldstone bosons.  By examining the size
of corrections to the one-loop effective action, Sundrum and Soldate
\ref\SS{M.~Soldate and R.~Sundrum, \np {B340} {1990} 1} have argued that
$\Lambda_\chi$ is bounded by $4\pi f/\sqrt{N}$.  We address this question by
examining pion scattering in chiral perturbation theory, and explicitly
confirm their results.  This result does not depend on any expansion in $1/N$.
In particular, if $N$ is significantly larger than 2 ({\it e.g.} 8 in the
one-family model), the chiral Lagrangian ceases to be valid at energies much
lower than suggested by naive dimensional analysis.  This change is
particularly important in the context of electroweak symmetry breaking
phenomenology at the SSC and LHC \ref\others{K.  Lane, in {\it Proceedings of
the 1982 DPF Summer Study on Elementary Particle Physics and Future
Facilities}, edited by R. Donaldson \etal, (Fermilab, Batavia, Ill.), p.
222\semi E. Eichten, I. Hinchliffe, G.  Lane, and C. Quigg, Rev.  Mod.  Phys.
{\bf 56} (1984) 579.} \baggerherr \ref\us{R. S. Chivukula and M.  Golden, \pl
{267B} {1991} {233}\semi R. S.  Chivukula, M. Golden, and M. V.  Ramana, \prl
{68} {1992} {2883}.}.

As previously stated, the chiral symmetries of technicolor are generally only
approximate symmetries. In addition to the three ``eaten'', exact Goldstone
bosons, there are often additional pseudo-Goldstone bosons.  If the mass of a
typical pseudo-Goldstone boson is $m$, chiral perturbation theory is also an
expansion in $m^2/\Lambda^2_\chi$, {\it i.e.} $\Lambda_\chi$ also sets the
scale for the size of corrections due to chiral symmetry breaking. Chiral
symmetry breaking interactions induce nonderivative terms in the chiral
Lagrangian.  For simplicity, in this paper we consider a chiral symmetry
breaking interaction which gives the same mass to all Goldstone bosons.  At
lowest order in the symmetry breaking the Lagrangian is
\eqn\lagbr{
\Lt = \Lt^{(0)} + {m^2 f^2 \over 4} \tr (\Sigma + \Sigma^\dagger)~.
}
We will show that in a theory with a large number of pseudo-Goldstone bosons,
since $\Lambda_\chi$ is smaller than expected, the effects of chiral symmetry
breaking are correspondingly larger than expected. While this particular kind
of symmetry breaking cannot occur in a technicolor theory -- such a symmetry
breaking term would break the weak gauge symmetry -- the calculation will
illustrate the enhancement of symmetry breaking effects in a theory with
$N > 2$.

In the next section we consider Goldstone boson scattering at tree level using
$\Lt$.  Using crossing invariance we show that all amplitudes may be expressed
in terms of two invariant functions. Decomposing the amplitude into flavor and
angular momentum channels, we find that the $SU(N)_V$ singlet, spin zero
channel would violate unitarity at an energy of order $4 \pi f/\sqrt{N}$.  As
physical amplitudes cannot violate unitarity, we can conclude that that terms
of order $p^4$ and higher must become important and that $\Lambda_\chi$ cannot
be larger than of order $4\pi f/\sqrt{N}$.  This argument is a generalization
of that applied by Lee, Quigg, and Thacker \ref\lqt{B. Lee, C. Quigg, and H.
Thacker, \prl {38} {1977} {883}.} to longitudinal gauge boson scattering in
the standard one Higgs-doublet model.

In the third section, we compute the corrections to Goldstone boson
scattering at one-loop. Confirming ref. \SS, we find that the corrections
are larger than naively expected by a factor of $N$. We give a simple
argument that this pattern persists to all orders in the loop expansion
implying again that $\Lambda_\chi$ is of order $4\pi f/\sqrt{N}$.
This implies that {\it all} terms in the chiral expansion are relevant
at these energies and chiral perturbation theory breaks down.

While the singlet channel saturates unitarity at energies of order $4 \pi
f/\sqrt{N}$, the other channels have amplitudes which are still much less than
one. Is it possible that the low energy predictions from $\Lt$ continue to be
valid in these other channels, as in a generalization of the ``conservative''
model of Chanowitz and Gaillard in \chanowitz?  In the fourth section, we
argue that in general, the chiral Lagrangian will be valid at all energies
below the energy scale associated with new physical states, \ie\
$\Lambda_\chi$ may always be interpreted as the scale of new physics and is
not just a formal artifact of chiral perturbation theory.  This new physics
will enter in all channels and hence the low-energy predictions for {\it all}
channels fail.  The behavior of the theory above $\Lambda_\chi$ depends on
what new physics is present, and one cannot trust arbitrary unitarizations of
the low-energy amplitudes (such as summing a subset of the chiral loop
diagrams, the $K$-matrix, or the Pad{\'e} approximants) \ref\cs{R.~N.~Cahn and
M.~Suzuki, \prl {67} {1991} {169}.}\ref\unitarize{K.  Jhung and R. Willey,
\prd {9} {1974} {3132}\semi A. Dobado and M. Herrero, \pl {228B} {1989}
{495}\semi A. Dobado, M. Herrero, and T.  Truong, \pl {235B} {1990} {129} and
134\semi A. Dobado, \pl {237B} {1990} {457}.}.

In QCD, the chiral Lagrangian ceases to be valid at an energy scale of order
the mass of the $\rho$ meson.  The arguments we make here imply that in a
theory with more than two light flavors, at least some of the resonances are
lighter than would be expected by scaling from QCD, and that any results for
such a theory based on scaling from QCD are suspect. In particular, the
estimates of ``oblique'' corrections in the one-family technicolor model
\ref\pt{M. Peskin, T. Takeuchi, \prl {65} {1990} {964}, and SLAC preprint
SLAC-PUB-5618 (Nov 1991)\semi R. N. Cahn and M. Suzuki, \prd {44} {1991}
{3641}.} are not trustworthy.

\newsec{Pion Scattering from $\Lt$}

Consider the scattering process $\pi^a \pi^b \to \pi^c \pi^d$.  The amplitude
for such a process may be decomposed into irreducible $SU(N)_V$
representations.  We may understand the representations contained in {\bf
adjoint} $\otimes$ {\bf adjoint} of $SU(N)$ as follows. Consider the object
$(T^a)^i_j (T^b)^k_l$, where the indices $a, b$ are in the {\bf adjoint} of
$SU(N)$ while $i,j,k,l$ are in the fundamental. First, we may trace $i$ with
$l$, and $j$ with $k$, yielding a singlet. Next, we may form an {\bf adjoint}:
remove the singlet already constructed, then symmetrize $(i,k)$,
antisymmetrize $(j,l)$, and contract $i$ with $l$. Another {\bf adjoint} is
obtained by symmetrizing both $(i,k)$ and $(j,l)$ before contracting. The next
possibility is to antisymmetrize $(i,k)$ and $(j,l)$, and remove all traces.
This representation has dimension $(N+1) (N-3) N^2 / 4$. We may symmetrize
$(i,k)$ and $(j,l)$ and remove traces, yielding a representation of dimension
$(N-1) (N+3) N^2 / 4$. Lastly there are two complex representations,
conjugates of each other, in which the traces are removed and either $(i,k)$
is symmetric and $(j,l)$ is antisymmetric or vice-versa.  These complex
representations have dimension $(N+2)(N+1)(N-1)(N-2)/4$. We refer to the seven
representations as $\Delta$, $F$, $D$, $Y$, $X$, $T$, and $\bar T$,
respectively. The representations $\Delta$, $D$, $Y$, and $X$ are symmetric
under the exchange of $a$ and $b$, and therefore only the angular momentum $J
= \hbox{even}$ partial waves can contribute. The others are antisymmetric
under $a \leftrightarrow b$, and only the odd angular momentum partial waves
contribute. For $N=2$ only $\Delta$, $F$, and $X$ exist, corresponding to the
isospin 0, 1, and 2 channels. If $N=3$, $Y$ does not exist, while the others
are respectively the $\bf 1, 8_a, 8_s, 27, 10,$ and $\bf \bar{10}$.

Next, we construct the most general amplitude for $\pi\pi$ scattering
consistent with Bose symmetry, crossing invariance, and $SU(N)_V$
conservation.  There are nine invariant tensors with four adjoint indices,
corresponding to the nine singlets in $({\bf adjoint})^4$.  A set of
nine\foot{There are nine when $N>3$.  Using the relations in Appendix A, all
the other invariants can be shown to dependent on these nine.  For $N=2$ the
$d$ symbols do not exist and there are only three invariants.  For $N=3$ the
relation (Burgoyne's identity \ref\sidney{S. Coleman, {\it Aspects of
Symmetry},(Cambridge University Press, 1985) p21}) $3(d^{abe}d^{cde} +
d^{ace}d^{bde} + d^{ade}d^{bce}) = \delta^{ab}\delta^{cd} +
\delta^{ac}\delta^{bd} + \delta^{ad}\delta^{bc}$ permits us to eliminate one
of the invariants.  } linearly independent invariants is
$\delta^{ab}\delta^{cd}$, $\delta^{ac}\delta^{bd}$, $\delta^{ad}\delta^{bc}$,
$d^{abe}d^{cde}$, $d^{ace}d^{bde}$, $d^{ade}d^{bce}$, $d^{abe}f^{cde}$,
$d^{ace}f^{bde}$, and $d^{ade}f^{bce}$, where $d^{abc}$ and $f^{abc}$ are
defined by
\eqn\fdef{
f^{abc} = -2i \tr [T^a,T^b] T^c \hbox{\hskip .2in and \hskip .2in}
d^{abc} =  2  \tr \{T^a,T^b\} T^c
{}~.}
There can be no part of the amplitude proportional to $d^{abe}f^{cde}$:
because of Bose symmetry and the symmetry of $d^{abc}$ under $a
\leftrightarrow b$, the initial state would have to be in an even angular
momentum state, while by the antisymmetry of $f^{cde}$, the final state would
have to be in an odd angular momentum state.  Therefore, the most general
amplitude is
\eqn\smatrix{
\eqalign{
a(s,t,u)^{a,b;c,d} =&
   \delta^{ab}\delta^{cd} A(s,t,u)
 + \delta^{ac}\delta^{bd} A(t,s,u)
 + \delta^{ad}\delta^{bc} A(u,t,s)\cr
&+ d^{abe} d^{cde} B(s,t,u)
 + d^{ace} d^{bde} B(t,s,u)
 + d^{ade} d^{bce} B(u,t,s)~,
}
}
where $s$, $t$, and $u$ are the Mandelstam variables and $A$ and $B$ are
unknown functions.  Bose symmetry also implies that the functions $A$ and $B$
must be symmetric under the exchange of their second and third arguments.

Applying the projection operators defined in Appendix A to the amplitude
\smatrix, the amplitudes for pion scattering in the various $SU(N)_V$ channels
can all be written in terms of $A$ and $B$:
\eqn\aAB{
\eqalign{
a_\Delta(s,t,u) =& (N^2-1)A(s,t,u)+A(t,s,u)+A(u,t,s) \cr
	&+ {(N^2-4)\over N}(B(t,s,u)+B(u,t,s))\cr
a_F(s,t,u) =& A(t,s,u)-A(u,t,s) + {N^2-4\over 2N}(B(t,s,u)-B(u,s,t))\cr
a_D(s,t,u) =& A(t,s,u)+A(u,t,s) + {N^2-4 \over N}B(s,t,u)\cr
	&+ {N^2-12\over 2N}(B(t,s,u)+B(u,t,s))\cr
a_Y(s,t,u) =& A(t,s,u)+A(u,t,s)-{N+2\over N}(B(t,s,u)+B(u,t,s))\cr
a_X(s,t,u) =& A(t,s,u)+A(u,t,s)+{N-2\over N}(B(t,s,u)+B(u,t,s))\cr
a_T(s,t,u) =& a_{\bar T}(s,t,u) = A(t,s,u)-A(u,t,s)-
  {2 \over N}(B(t,s,u)-B(u,t,s))\cr
}
}
The lowest order chiral Lagrangian gives $A(s,t,u)=(2/N)(s-m^2)/f^2$, and
$B(s,t,u)=(s-m^2)/f^2$.

The partial wave amplitudes are defined by
\eqn\wavedef{
a_{I\ell}(s) = {1\over 64 \pi} \int_{-1}^1 a_I(s, \cos\theta)
P_\ell(\cos\theta) d\cos\theta~,
}
where $P_\ell$ is the Legendre polynomial of order $\ell$.
The scattering amplitudes in the various channels from
$\Lt$ \cs are then:
\eqn\scatt{
\eqalign{
a_{\Delta 0} &= {N s \over 32 \pi f^2} - {m^2 \over 16 \pi N f^2} \cr
a_{F 1} &= {N s \over 192 \pi f^2} - {N m^2 \over 48 \pi f^2}\cr
a_{D 0} &= {N s \over 64 \pi f^2} - {m^2 \over 8 \pi N}\cr
a_{Y 0} &= {s \over 32 \pi f^2} - {m^2 \over 16 \pi f^2}\cr
a_{X 0} &= -{s \over 32 \pi f^2} + {m^2 \over 16 \pi f^2}~.
}
}
All other partial wave amplitudes are zero (including $a_{T 1}$ and
$a_{\bar{T} 1})$. Note that $a_{\Delta 0}$, $a_{F 1}$ and $a_{D 0}$
are enhanced by a factor of $N$.

The amplitude $a_{\Delta 0}$ calculated at tree level is real, and (for small
$m^2$) would exceed 1 when $\sqrt{s} > 4 \pi f / \sqrt{N}$.  A physical
scattering amplitude must lie on or inside the Argand circle.  At these
energies, therefore, loop corrections and higher order terms in the chiral
Lagrangian must make as large a contribution as the two-derivative term, and
the calculation using $\Lt$ ceases to be useful. This suggests that
$\Lambda_\chi$ is less than or of order $4 \pi f/\sqrt{N}$, as was emphasized
in \SS.  Note that this result holds independent of the largeness of $N$.  No
expansion in powers of $1/N$ need be made.

\newsec{Pion Scattering at Order $p^4$}

An alternative approach to put a limit $\Lambda_\chi$ is based on an estimate
of the size of loop corrections \wein.  Since the theory is not
renormalizable, the terms of order $p^4$ are required as counterterms to loops
involving the lowest order interactions.  In calculating the scattering
amplitude to order $p^4$, one must consider tree-level diagrams with
interactions coming from operators of fourth order in momenta, and one-loop
diagrams using $\Lt$. It is unnatural to assume that the contribution from the
former is much larger than the latter since such a statement could only be
true for a particular choice of renormalization scale. Similarly, the two-loop
calculation using $\Lt$ will require counterterms of order $p^6$, etc.  In
this section we compute the one-loop corrections to Goldstone boson
scattering.

The next-to-leading order Lagrangian is made up of terms containing four
derivatives, two derivatives and one power of the symmetry breaking parameter,
or two powers of the symmetry breaking:
\eqn\lagfour{
\eqalign{
\Lf =&
 \ell_1  \{\tr(\partial^\mu\Sigma^\dagger\partial_\mu\Sigma)\}^2 +
 \ell_2  \{\tr(\partial^\mu\Sigma^\dagger\partial^\nu\Sigma)\}^2 +\cr
&\ell_3  \tr(\partial^\mu\Sigma^\dagger\partial_\mu\Sigma
               \partial^\nu\Sigma^\dagger\partial_\nu\Sigma) +
 \ell_3' \tr(\partial^\mu\Sigma^\dagger\partial^\nu\Sigma
               \partial_\mu\Sigma^\dagger\partial_\nu\Sigma) + \cr
&\ell_4  m^2\tr(\Sigma + \Sigma^\dagger)
              \tr(\partial^\mu\Sigma^\dagger\partial_\mu\Sigma) +
 \ell_5  m^2\tr\{(\Sigma + \Sigma^\dagger)
                  (\partial^\mu\Sigma^\dagger\partial_\mu\Sigma)\} + \cr
&\ell_6  m^4\{\tr(\Sigma + \Sigma^\dagger)\}^2 +
 \ell_7  m^4\{\tr(\Sigma - \Sigma^\dagger)\}^2 +
 \ell_8  m^4\tr(\Sigma^2 + \Sigma^{\dagger2})~.
}
}
All other possible terms vanish by the equations of motion.  This notation
agrees with that of Gasser and Leutwyler\ref\Gasser{Gasser and Leutwyler, Ann.
Phys. (NY) {\bf 158} (1984) 142 and \np {B250} {1985} {465}.}.
In their case the term proportional to $\ell_3'$ is not linearly independent,
because they were considering an $SU(3)\times SU(3)$ chiral symmetry.

The calculation of the $\pi \pi \to \pi \pi$ amplitudes at one loop is
straightforward, if tedious.  The result is:
\eqn\loopresa{
\eqalign{
A(&s,t,u) = {2\over N} {s - m^2\over f^2}\cr
&+{J(s)\over f^4}\left(-{1\over 2}s^2+{2\over N^2}m^4\right)
 +{J(t)\over f^4}\left(
  -{1\over3}t^2+{5\over3}m^2t-{7\over3}m^4-{1\over6}st+{2\over3}m^2s\right)
 \cr
&+{J(u)\over f^4}\left(
  -{1\over3}u^2+{5\over3}m^2u-{7\over3}m^4-{1\over6}su+{2\over3}m^2s\right)
 \cr
&+ {s^2 \over f^4}
  \left(-{2\over3}\DD+8\ell_1(\mu)+4\ell_2(\mu)+{8\over N}\ell_3(\mu)
  +{23\over18}\spi\right)
 \cr
&+ {tu \over f^4}
  \left({2\over3}\DD-8\ell_2(\mu)-{16\over N}\ell_3'(\mu)-{13\over9}\spi\right)
 \cr
&+ {m^2s \over f^4}
  \biggl({2\over3}\DD-32\ell_1(\mu)-16\ell_2(\mu)-{32\over N}\ell_3(\mu)
 \cr
&\hbox{\hskip .2in}+16\ell_4(\mu) + {16\over N}\ell_5(\mu)
  -{28\over9}\spi\biggr)
 \cr
&+ {m^4 \over f^4}
 \biggl(\left(- 2 + {2\over N^2}\right)\DD+32\ell_1(\mu)+32\ell_2(\mu)
  +{32\over N}\ell_3(\mu)+{32\over N}\ell_3'(\mu)
  \cr
&\hbox{\hskip .2in}-32\ell_4(\mu)-{32\over N}\ell_5(\mu)+32\ell_6(\mu)
  +{32\over N}\ell_8(\mu) + \left({20\over3} - {4\over N^2}\right)\spi\biggr)
  \cr}}
\eqn\loopresb{
\eqalign{
B(&s,t,u) = {s - m^2\over f^2}\cr
&+{NJ(s) \over f^4}\left(-{1\over8}s^2+{2\over N^2}m^4\right)
 +{NJ(t) \over f^4}\left(
  -{1\over24}t^2+{1\over3}m^2t-{2\over3}m^4-{1\over12}st+{1\over3}m^2s\right)
  \cr
&+{NJ(u) \over f^4}\left(
  -{1\over24}u^2+{1\over3}m^2u-{2\over3}m^4-{1\over12}su+{1\over3}m^2s\right)
  \cr
&+ {Ns^2 \over f^4}
 \left( - {1\over12}\DD+{4\over N}\ell_1(\mu)+{5\over36}\spi\right)
 \cr
&+ {Ntu \over f^4}
 \left(+{1\over12}\DD-{8\over N}\ell_3'(\mu)-{2\over9}\spi\right)
 \cr
&+ {Nm^2s \over f^4}
 \left(- {1\over6}\DD-{16\over N}\ell_3(\mu)+{8\over N}\ell_5(\mu)
  -{5\over9}\spi\right)
 \cr
&+ {N m^4 \over f^4}
 \biggl({2\over N^2}\DD+{16\over N}\ell_3(\mu)+{16\over N}\ell_3'(\mu)
 -{16\over N}\ell_5(\mu)+{16\over N}\ell_8(\mu)
  \cr
&\hbox{\hskip .2in}+\left({4\over3}-{4\over N^2}\right)\spi\biggr)
}
}
Here we have used dimensional regularization in $4-\epsilon$ dimensions with a
scale $\mu$ and, as detailed in Appendix B, the parameters $\ell_i(\mu)$ are
renormalized at this scale.  The function $J$ is defined as
\eqn\Jdef{
J(x) = {1\over 16\pi^2} \sqrt{1 - {4 m^2\over x}}
\log {\sqrt{1 - {4 m^2\over x}}+1 \over \sqrt{1 - {4 m^2\over x}}-1}
}
where $\log(z)$ and $\sqrt{z}$ are both understood to have a cut under the
negative real axis.  The quantities $m$ and $f$ which appear in \loopresa\ and
\loopresb\ are the physical mass and decay constant.  Their renormalizations
are also given in Appendix B.

 From \loopresa\ and \loopresb\ we see that the typical corrections to
the tree level results are of order $N s/(16 \pi^2 f^2)$ or $N m^2/(16 \pi^2
f^2)$. In agreement with the previous section and ref. \SS, then, it is
inconsistent to assume that $\Lambda_\chi$ is much greater than $4\pi
f/\sqrt{N}$.

This pattern, {\it i.e.} that loop results are larger by powers of $N$ than
expected by naive dimensional analysis, persists to all orders. In order to
keep track of $N$, we may use a double line notation for each Goldstone boson,
analogous to the double line notation used by 't Hooft \ref\thooft{G. 't
Hooft, \np {B72} {1974} {461}.} in large-$N_C$ QCD. Each Goldstone boson is a
member of the adjoint representation of $SU(N)_V$, and is analogous to a gluon
in the adjoint of the $SU(N_C)$ color.  Analogous to large-$N_C$ QCD, then,
the contributions to a process at $L$ loops using interactions from $\Lt$
that have the highest power of $N$ are proportional to $N^L/(16 \pi^2)^L$.
Therefore, there are $L$-loop corrections to pion scattering of order
\eqn\typical{
\left( \sqrt{N} p \over 4 \pi f \right)^{2L}~,
}
where $p$ is a typical momentum in the process.  Since all higher momentum
corrections are suppressed by the same energy scale, chiral perturbation
theory as a whole breaks down by energies of order $4 \pi f/\sqrt{N}$.  For
any fixed $N$, it is {\it possible} that the subleading in $N$ diagrams are
numerically as important as those that are leading.  However, making
$\Lambda_\chi$ much bigger than $4 \pi f / \sqrt{N}$ would require both an
enhancement at every loop of the subleading diagrams by a numerical factor as
large as $N$, and a cancellation of these with the leading diagrams.

\newsec{Implications of Analyticity and Crossing}

Some interesting questions arise at this point.  We have argued that chiral
perturbation theory breaks down at or before $\Lambda_\chi$, but what actually
happens to the amplitudes as $s$ increases beyond this value? What is the
significance of $\Lambda_\chi$?  The amplitudes for the partial waves other
than $a_{\Delta 0}$ are all below their unitarity limits when $\sqrt{s} = 4
\pi f / \sqrt{N}$.  Is it possible that these other channels continue to
behave like the prediction of the lowest order chiral Lagrangian, as in the
$SU(N)_V$ generalization of the ``conservative model'' of ref.  \chanowitz?

The pion scattering amplitudes are determined by the two functions $A$ and
$B$.  We will examine the analytic structure of these functions in the complex
$s$, $t$ hypersurface, where $u$ is determined by the mass shell condition.
Let us consider the case of a small $m^2>0$, so as to avoid the subtleties
associated with infrared problems.  Amplitudes in chiral perturbation theory
are expansions in $s$ and $t$.  The $S-$matrix is analytic on this
hypersurface except for cuts on the physical $(s,t)$ plane and poles or cuts
on the unphysical hypersheets.  The cuts in the physical $(s,t)$ plane are due
to multipion states, and the appearance of poles or other structure on the
unphysical hypersheets correspond to resonances or physical states other than
pions.  The masses of the new physical states are not protected by a chiral
symmetry and the corresponding structures in the $S$-matrix are therefore
away from the origin.

Consider the region $R$ with $s$, $t$, and $u$ all less than $4m^2$. Here,
assuming there are no particles lighter than twice the pion mass, the
$S$-matrix is analytic.  The functions $A$ and $B$ computed from the chiral
Lagrangian to arbitrary order have an infinite set of adjustable coefficients
multiplying the terms $s^it^j$ for all $i$ and $j$.  We can regard this as a
convergent expansion about any point in the region $R$.  Therefore, the
amplitude computed in chiral perturbation theory can be adjusted to match the
$S-$matrix exactly in $R$. Going outside $R$, the singularities of the
$S$-matrix on the physical plane correspond to cuts from multipion states and
are determined by unitarity.  They are correctly included in chiral loop
calculations.  Moving away from this region, therefore, the chiral Lagrangian
calculation should reproduce the $S$-matrix so long as there aren't any
singularities closer to the origin associated with physical states {\it other}
than multipion states. That is, chiral perturbation theory is a good
approximation to the $S$-matrix {\it all the way} out to an energy scale
associated with the appearance of new physics.

Admittedly, the argument given above for the range of validity of the chiral
expansion is not truly rigorous.  This is because it works only if one
computes without expanding in $m^2$. As emphasized by Pagels and Li
\ref\pagels{L.-F. Li and H. Pagels, \prl {26} {1971} 1089\semi see also H.
Pagels, Phys. Reps. {\bf 16C} (1975) 221.}, the $S$-matrix is not analytic in
$m^2$. However, as motivated by the theory of critical phenomena \ref\ma{See,
for example, Shang-Keng Ma, {\it Modern Theory of Critical Phenomena}
(Benjamin/Cummings, 1976)}, we assume that the coefficients of the Chiral
Lagrangian are analytic in $m^2$ and that the nonanalyticity of the $S$-matrix
is reproduced by chiral loop calculations.

If this standard assumption is correct, then the arguments given above are
correct so long as $m^2$ is small enough, and the chiral expansion is valid
for all energies below the first nonanalytic structure in the $S-$matrix
corresponding to some real new physics. The breakdown of the chiral Lagrangian
is not a calculational artifact; the scale $\Lambda_\chi$ has direct physical
significance, unlike the scale $\Lambda_{QCD}$ in perturbation theory.
Moreover, the arguments above show that this physical scale cannot be larger
than about $4\pi f / \sqrt{N}$.

In general, there are many possibilities for the new physics at
$\Lambda_\chi$.  In QCD with $N=2$, the $\rho$ pole gives a singularity in $A$
when $t$ or $u$ is $m_\rho^2 - i m_\rho \Gamma_\rho$.  In the one-doublet
Higgs model, there is a pole in $A$ whenever $s$ is $M_H^2 - i M_H \Gamma_H$.
The first singularity to appear could be a branch cut instead of a pole.
Consider a world in which the $\pi$s are massless, but the $K$s weigh 200 MeV.
(The pion decay constant is still 93 MeV.)  Imagine constructing an $SU(2)$
invariant chiral Lagrangian to describe the massless pions.  The first new
singular structure one encounters in the $S$-matrix is the two $K$ branch cut,
at 400 MeV, and at this energy the chiral expansion breaks down.

Since the scale $\Lambda_\chi$ is associated with non-universal structure,
such as the $\rho$, the Higgs, or the $KK$ states in the examples above, the
behavior of the scattering amplitudes at energies at or above $\Lambda_\chi$
{\it cannot be} universal. Therefore, there is no reason to trust predictions
based on an arbitrary, {\it e.g.} $K$-matrix, Pad{\'e} approximant\unitarize,
or bubble sum\cs, unitarization of the universal, lowest-order, chiral
amplitudes.

Finally, we address the issue of whether the channels other than $a_{\Delta
0}$ can follow their low-energy predictions beyond energies of order
$\Lambda_\chi$.  Consider again the functions $A$ and $B$.  Because of the
crossing relations \aAB, all channels will be affected by whatever new physics
enters at the scale $\Lambda_\chi$.  In particular we expect that at
$\Lambda_\chi$ the amplitudes in all channels will deviate strongly from the
low-energy predictions, even though the low-energy predictions may be well
below 1.

\nfig\aoo{Data and lowest-order prediction for $Re\ a_{\Delta 0}$.
The data is from a compilation in ref. \drv\ of the experimental results
in ref. \dataref.}
\nfig\aii{Data and lowest-order prediction as in \aoo, for $|a_{F1}|$.}
\nfig\ato{Data and lowest-order prediction as in \aoo, $Re\ a_{X0}$.}

Consider pion scattering in QCD. The data\foot{We thank G.  Valencia for
providing us with the data compiled for reference \ref\drv{J.~F.~Donoghue,
C.~Ramirez, and G.~Valencia, \prd {38} {1988} {2195}.}.} and lowest-order
predictions for the low isospin channels are plotted in \figs{\aoo, \aii,
\ato}.  The amplitude in the isospin-0 spin-0 channel ($a_{\Delta 0}$) starts
to deviate from the lowest order prediction at an energy of about 600 MeV.
This is not surprising, because the amplitude is at that point a substantial
fraction of its unitarity limit. Examining the other partial waves, we find
that they deviate from the lowest order chiral Lagrangian formulas above
approximately {\it the same energy scale}.  For example, in the isospin-1
spin-1 channel ($a_{F 1}$), the $\rho$ resonance appears at 770 MeV. While we
cannot predict the appearance of the $\rho$, the fact that the isospin 1 and 2
amplitudes deviate strongly from their low-energy predictions while they would
still be relatively weakly coupled is not a surprise.

\newsec{Conclusions and Speculations}

In this paper, we have argued that in general, the chiral expansion will be
valid at all energies below the energy scale associated with new physical
states.  That is, $\Lambda_\chi$ may always be interpreted as the scale of new
physics and is not just a formal artifact of chiral perturbation theory.  In
general, there is no procedure to infer the high-energy structure of the
theory from the universal lowest-order chiral lagrangian.  In the three
examples given in section four, the high-energy physics is different, while
the low-energy behavior is precisely the same.  In particular, there is no
reason to trust the amplitudes constructed by unitarizing the lowest-order
predictions.

In a theory with a spontaneously broken $SU(N)_L \times SU(N)_R$ chiral
symmetry, the scale $\Lambda_\chi$ is bounded by $4\pi f/\sqrt{N}$.  This
result depends only on the low-energy effective theory and is independent of
the precise form of the fundamental theory.  If the theory is QCD-like,
$\Lambda_\chi$ is presumably associated with the appearance of resonances like
the $\rho$. When $N=2$ and $f_\pi = 93$ MeV, the limit on $\Lambda_\chi$ is
about 825 MeV, and the $\rho$ is close to saturating this bound.

In QCD with $N$ flavors the masses of the vector mesons must decrease relative
to $f_\pi$ at least as fast as $1/\sqrt{N}$ as the number of flavors
increases. This behavior is illustrated in one limit. Consider QCD with $N$
flavors and $N_C$-colors in the limit $N_C \to \infty$ and $N \to \infty$,
with $N/N_C$ held fixed. In the fundamental theory at $L$ loops, the largest
contributions to any Green's function are proportional to $g^{2L} N^{L-k}_C
N^k$ for $0 \leq k \leq L$ and where $g$ is the strong coupling constant.  As
in conventional large-$N_C$ QCD \thooft\ this theory is well-behaved if we
hold $g^2 N_C$ fixed. In this limit, the masses of resonances are fixed and
$f_\pi$ grows like $\sqrt{N_C}$ or, equivalently, $\sqrt{N}$. Therefore, the
ratio of the masses of resonances to $f_\pi$ falls like $1/\sqrt{N}$.

We wish to stress, however, that the behavior of $\Lambda_\chi$ with $N$ is
{\it not} dependent on taking the double limit.  For example, when $N_C = 3$,
we know that QCD is well defined for the six quarks that actually exist.  If
they were all light and $f_\pi$ were fixed at 93 MeV, the masses of the
analogs of the $\rho$ would have to be less than 500 MeV!

In the one-family technicolor model, $N=8$ and $f \approx 125$ GeV. The low
energy prediction for the singlet spin zero channel would saturate unitarity
at an energy of order 550 GeV. The arguments we have given imply that the
resonances of this model must have masses of this order of magnitude.  In
technicolor phenomenology, however, one usually scales from QCD \ref\dkr{S.
Dimopoulos, S. Raby, and G. Kane, \np {B182} {1981} {77}} by multiplying all
mass scales by $f_{TC}/f_\pi$ and applying large $N_{TC}$ arguments \thooft.
When $N_{TC} = 3$ in the one-family model, this gives a technirho mass of
about 1000 GeV.  While we cannot say conclusively that there are {\it
resonances} as light as 550 GeV, it is clear that the one-family model is
quite different from QCD and that predictions based on scaling from QCD, e.g.
\others, cannot be trusted in detail.  In particular, if the masses of the
resonances are much lighter than naively expected, then the estimates of
``oblique'' corrections from technicolor obtained by scaling from QCD \pt, are
untrustworthy.

\appendix{A} {Projection Operators and Identities for $SU(N)$}

The following relations may be derived from the $SU(N)$ Fierz identities:
\eqn\relations{
\eqalign{
f^{abe}f^{cde} =& d^{ace}d^{bde} - d^{ade}d^{bce} + {2 \over N}
(\delta^{ac}\delta^{bd} - \delta^{ad}\delta^{bc}) \cr
f^{abe}d^{cde} =& d^{ade}f^{bce} + d^{ace}f^{bde}\cr
d^{aef} d^{bef} =& {N^2 - 4 \over N} \delta^{ab}\cr
d^{aef} d^{bfg} d^{cge} =& {N^2 - 12 \over 2N} d^{abc}\cr
d^{aef}d^{bfg}d^{cgh}d^{dhe} =& {N^2-4 \over N^2}
(\delta^{ab}\delta^{cd} + \delta^{ad}\delta^{bc})
+ {N^2 - 16 \over 4 N} (d^{abe}d^{cde} + d^{ade}d^{bce}) \cr
&-{N \over 4} d^{ace}d^{bde} \cr
}
}

The operators that project a state $|\pi^a \pi^b\rangle$ onto the various
irreducible representations of $SU(N)_V$ are
\eqn\Pdef{
\eqalign{
P_\Delta =& {1 \over N^2-1} \delta^{ab}\delta^{cd}\cr
P_F =& {1 \over N} (d^{ace}d^{bde} - d^{ade}d^{bce}) + {2 \over N^2}
  (\delta^{ac}\delta^{bd} - \delta^{ad}\delta^{bc})\cr
P_D =& {N \over N^2-4} d^{abe}d^{cde}\cr
P_Y =& {N-2 \over 4 N} (\delta^{ac} \delta^{bd} + \delta^{ad}\delta^{bc})
  + {N-2 \over 2 N (N-1)} \delta^{ab}\delta^{cd}\cr
 &- {1\over 4} (d^{ace}d^{bde} + d^{ade}d^{bce})
  + {N-4 \over 4(N-2)} d^{abe}d^{cde}\cr
P_X =& {N+2 \over 4 N} (\delta^{ac} \delta^{bd} + \delta^{ad}\delta^{bc})
  - {N+2 \over 2 N (N+1)} \delta^{ab}\delta^{cd}\cr
 &+ {1\over 4} (d^{ace}d^{bde} + d^{ade}d^{bce})
  - {N+4 \over 4(N+2)} d^{abe}d^{cde}\cr
P_T = P_{\bar T}^* =& {N^2-4 \over 4N^2}
  (\delta^{ac}\delta^{bd}-\delta^{ad}\delta^{bc})\cr
 &- {1 \over 2N} (d^{ace}d^{bde}-d^{ade}d^{bce})
  + {i \over 4} (d^{ace}f^{bde}-d^{ade}f^{bce}) ~.\cr
}
}
Using the identities \relations, one can verify that these are projection
operators satisfying
\eqn\addone{P_I^{a,b;c,d}P_K^{c,d;g,h} = \delta_{IK} P_I^{a,b;g,h},}
and
\eqn\addtwo{\Sigma_I P_I^{a,b;c,d} = \delta^{ac} \delta^{bd}~.}

We may write the amplitude in terms of the seven $SU(N)_V$
channels:
\eqn\smatrixII{
a(s,t,u)^{a,b;c,d} = \sum_I a_I(s,t,u) P_I^{a,b;c,d}~,
}
where $I$ runs over $\Delta, F, D, Y, X, T, \bar{T}$,
yielding eqn. \aAB.

\appendix{B} {Renormalization of $\Lf$}

Calculating at one loop, there are renormalizations to the pion mass, the
decay constant, and the pion wave function.  We define
\eqn\renorms{
\eqalign{
\pi^a &= Z_\pi^{-1/2} \pi^a_0\cr
m &= Z_m m_0\cr
f &= Z_f f_0\cr
}
}
where the subscript 0 denotes the bare quantities.  The calculation of the
mass and wavefunction renormalizations is straightforward.  We demand that the
pion propagator have a pole of residue 1 at $m^2$.  There is only one one-loop
diagram which contributes.  In addition some terms in ${\cal L}_4$ make a
contribution, and the total is
\eqn\renmpi{
\eqalign{
Z_\pi &= 1 - {m^2\over f^2} \left(8N\ell_4 + 8\ell_5 +
  {N \over 3} \left(D+\spi-\lgmm\right)\right)\cr
Z_m &= 1 - {m^2\over f^2} \left(4N\ell_4 + 4\ell_5 - 8N\ell_6 - 8\ell_8 +
  {1 \over 2N} \left(D+\spi-\lgmm\right)\right)\cr
}
}
Here $\mu$ is scale of dimensional regularization and and $D$ is the quantity
$\spi(2/\epsilon - \gamma_E + \log 4 \pi)$.

To renormalize $f$, we must define the axial current.  To do this we replace
$\partial^\mu \Sigma$ by $(\partial^\mu - i a^\mu) \Sigma - i \Sigma a^\mu$ in
$\cal L$.  The axial current $A^\mu$ is what multiplies $a^\mu$ in $\cal L$;
at lowest order it is $A_\mu^a = (if^2/2) \tr(T^a(\Sigma^\dagger \partial_\mu
\Sigma - \Sigma\partial_\mu \Sigma^\dagger))$. The decay constant $f$ is then
defined by
\eqn\pcac{
\langle 0 | A^{a\mu} | \pi^b \rangle = i f p^\mu \delta^{ab}
}
The operators multiplying $\ell_1\ldots\ell_3'$ make a contribution to
$A^\mu$, but not to the matrix element in \pcac.  The only contributions in
$\Lf$ are from the $\ell_4$ and $\ell_5$ terms.  In addition there is a
one-loop diagram from $\Lt$.  The result is
\eqn\renf{
Z_f = 1 + {m^2\over f^2} \left(4N\ell_4 + 4\ell_5 +
  {N \over 2} \left(D+\spi-\lgmm\right)\right)
}

Finally, the infinities in the computation of the scattering amplitudes can
all be absorbed by defining renormalized quantities $\ell(\mu)$ by
\eqn\renell{
\eqalign{
\ell_1(\mu) &=  \ell_1 + {1\over32} D \hbox{\hskip .2in}
\ell_2(\mu) =  \ell_2 + {1\over16} D \hbox{\hskip .2in}
\ell_3(\mu) =  \ell_3 + {N\over48} D \hbox{\hskip .2in}\cr
\ell_3'(\mu) &= \ell_3' + {N\over96} D \hbox{\hskip .2in}
\ell_4(\mu) =  \ell_4 + {1\over16} D \hbox{\hskip .2in}
\ell_5(\mu) =  \ell_5 + {N\over16} D \hbox{\hskip .2in}\cr
\ell_6(\mu) &=  \ell_6 + \left({1\over32}+{1\over16N^2}\right)
  D \hbox{\hskip .2in}
\ell_8(\mu) =  \ell_8 + \left({N\over32}-{1\over8N}\right)
  D \hbox{\hskip .2in}
}
}
This computation does not yield the renormalization of $\ell_7$ because that
term makes no contribution to $\pi\pi\to\pi\pi$ scattering.  Gasser and
Leutwyler have computed the finite parts of $A(s,t,u)$ for the case of $N=2$
and our formula agrees with theirs\foot{They do not use the $\overline{MS}$
prescription.  Their subtractions are proportional to $D + 1/16\pi^2$ instead
of $D$.}.  The results in \renell\ agree with their computation for the case
of $N=3$, after one eliminates the operator corresponding to $\ell_3'$.

\bigskip

\noindent {\bf Acknowledgements. }\hfil\break

We thank Andrew Cohen, Howard Georgi, Ken Lane, Stephen Selipsky, and
Elizabeth Simmons for useful conversations and German Valencia for providing
us with the data compiled in ref. \drv.  R.S.C.  acknowledges the support of a
NSF Presidential Young Investigator Award and of an Alfred P.  Sloan
Foundation Fellowship.  M.G.  thanks the Texas National Research Laboratory
Commission for support under a Superconducting Super Collider National
Fellowship. This research is supported in part by the National Science
Foundation under grants PHY--87--14654 and PHY--90--57173, by the Texas
National Research Laboratory Commission, under grants RGFY9106 and RGFY91B6,
and by the Department of Energy, under contract numbers DE--AC02--89ER40509
and DE--FG02--91ER40676.

\nref\dataref{
N. Cason, \etal, \prd {28} {1983} {1586}\semi
V. Srinivasan, \etal, \prd {12} {1975} {681}\semi
L. Rosselet, \etal, \prd {15} {1977} {574}\semi
A. Belkov, \etal, JETP Lett. 29 (1979) 597\semi
B. Hyams, \etal, \np {B73} {1974} {202}\semi
E. Alekseeva, \etal, Sov. Phys. JETP 55 (1982) 591 and JETP Lett. 29 (1979)
100 \semi
W. Hoogland, \etal, \np {B69} {1974} {266}\semi
J. Baton, \etal, \pl {33B} {1970} {525} and 528\semi
J. Prukop, \etal, \prd {10} {1974} {2055}\semi
D. Cohen, \etal, \prd {7} {1973} {662}.}

\listrefs
\listfigs

\bye